\documentclass{iau} 
\usepackage{graphicx}

\def \halpha{H$\alpha$}
\def \Msol{{\rm M}_{\odot}}
\def \Lsol{{\rm L}_{\odot}}

\def \logm{\log(M/\Msol)}

\def \lya{Ly$\alpha$}

\def \h2{{\rm H_{2}}}

\def \halpha{H$\alpha$~}

\def \Cii{[C{\scriptsize ~II}]}

\def \heii{He{\scriptsize ~II}}

\def \IRXB{IRX$-\beta$}
\def \dn4000{D_{{\rm n}}(4000) }

\title[Panchromatic Study of the First Galaxies] 
{Panchromatic Study of the First Galaxies with Large ALMA Programs}

\author[Andreas Faisst]   
{A. Faisst$^1$, M. B\'ethermin$^2$, P. Capak$^1$, P. Cassata$^3$, O. LeF\`evre$^2$, D. Schaerer$^4$, J. Silverman$^5$, L. Yan$^{1,6}$, and the \textit{ALPINE} team}

\affiliation{$^1$IPAC, California Institute of Technology, \\ 1200 East California Boulevard, Pasadena, CA 91125, USA \\ email: {\tt afaisst@ipac.caltech.edu} \\[\affilskip]
$^2$Laboratoire d'Astrophysique de Marseille \\ UMR 7326, 13388, Marseille, France\\[\affilskip]
$^3$University of Padova, Department of Physics and Astronomy \\ Vicolo Osservatorio 3, 35122, Padova, Italy\\[\affilskip]
$^4$Observatoire de Gen\`eve, Universit\'e de Gen\`eve \\ 51 Ch. des Maillettes, 1290 Versoix, Switzerland \\[\affilskip]
$^5$Kavli Institute for the Physics and Mathematics of the Universe, The University of Tokyo \\ Kashiwa, Chiba 277-8583, Japan \\[\affilskip]
$^6$Caltech Optical Observatories, Cahill Center for Astronomy
and Astrophysics \\ 1200 East California Boulevard, Pasadena, CA 91125, USA}

\pubyear{2018}
\volume{341}  
\jname{Challenges in Panchromatic Modelling\\ with Next Generation Facilities}
\editors{M. Boquien, E. Lusso, C. Gruppioni \& P. Tissera}
\begin{document}

\maketitle

\begin{abstract}
Thanks to deep optical to near-IR imaging and spectroscopy, significant progress is made in characterizing the rest-frame UV to optical properties of galaxies in the early universe ($z>4$). Surveys with Hubble, Spitzer, and ground-based facilities (Keck, Subaru, and VLT) provide spectroscopic and photometric redshifts, measurements of the spatial structure, stellar masses, and optical emission lines for large samples of galaxies.
Recently, the Atacama Large (Sub) Millimeter Array (ALMA) has become a major player in pushing studies of high redshift galaxies to far-infrared wavelengths, hence making panchromatic surveys over many orders of frequencies possible. While past studies focused mostly on bright sub-millimeter galaxies, the sensitivity of ALMA now enables surveys like \textit{ALPINE}, which focuses on measuring the gas and dust properties of a large sample of \textit{normal} main-sequence galaxies at $z>4$.
Combining observations across different wavelengths into a single, panchromatic picture of galaxy formation and evolution is currently and in the future an important focus of the astronomical community.

\end{abstract}

\firstsection 

\section{Overview}

\begin{figure}[t!]
\begin{center}
\includegraphics[width=0.94\columnwidth]{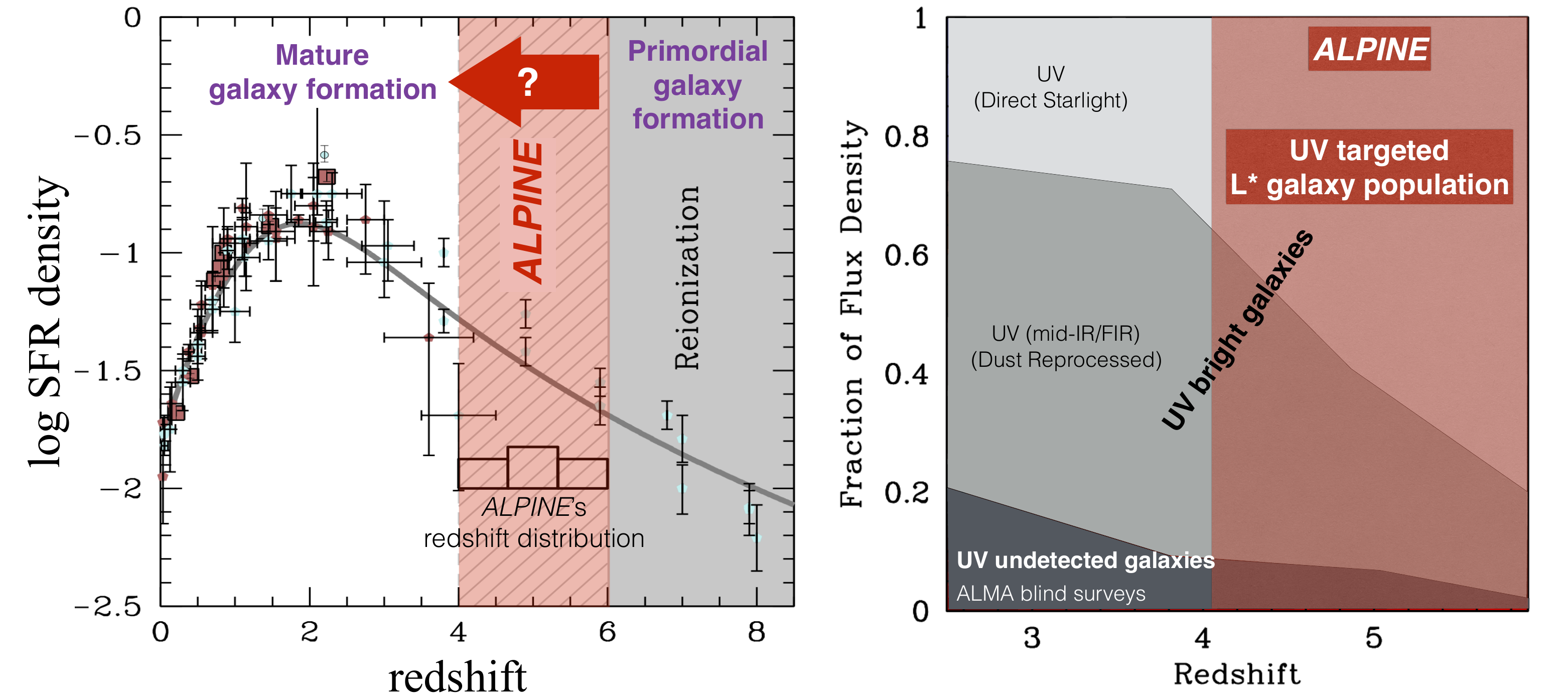}
\caption{\textit{Left:} SFR density across cosmic time (figure adapted from \cite[Madau \& Dickinson (2014)]{MADAU14}). \textit{ALPINE} targets galaxies in the early growth phase, a transition phase between primordial and mature galaxy formation where galaxies substantially increase their stellar mass, dust, and metal content.
\textit{Right:} Fraction of flux density emitted by galaxies directly in the UV (top shading), by UV-selected galaxies in the far-IR (middle shading), and by galaxies only probed in blind far-IR surveys (bottom shading). Figure adapted from \cite[Bouwens \etal\ (2009)]{BOUWENS09}.\label{fig:fig1}}
\end{center}
\end{figure}

Galaxies evolve significantly over cosmic time in both internal as well as external properties. The cosmic star formation rate (SFR) density of the early universe increases rapidly to a redshift $z=2$ (Figure~\ref{fig:fig1}) and galaxies in the early universe double their stellar mass at $>10$ times faster rates compared to today (\cite[Faisst et al. 2016a]{FAISST16a}). This steep mass growth goes in hand with a significant increase of the dust and metal content (\cite[Maiolino \etal\ 2008, Bouwens \etal\ 2009, Faisst \etal\ 2016b]{MAIOLINO08,BOUWENS09,FAISST16b}). 
In addition, star-forming galaxies evolve from lumpy and turbulent (dominated by gas inflow, mergers, and star-forming clumps) to smooth disks as cosmic time progresses (\cite[F{\"o}rster-Schreiber \etal\ 2009]{FORSTERSCHREIBER09}).

In only $500\,{\rm Myrs}$ between redshifts $4 < z < 6$, galaxies significantly increase their stellar, dust, and gas mass as well as metal content and in addition change their structure by building up the first stellar disks. Hence, this \textit{early growth phase} is an important link between primordial (during the Epoch of Reionization at $z>4$) and mature (at the peak of cosmic SFR density at $z\sim2-4$) galaxy formation.

To date, more than $1000$ galaxies have been identified at $4 < z < 6$ by photometric redshifts, color-color selections, and narrow-band observations targeting \lya. Several hundred galaxies have been confirmed spectroscopically.
Large surveys with HST specifically focus on the galaxies' UV properties including their luminosity function, UV dust obscuration, and morphology (e.g., \cite[Bouwens \etal\ 2015, Shibuya \etal\ 2015]{BOUWENS15,SHIBUYA15}). Observations with Spitzer at $3-5\,{\rm \mu m}$ (rest-frame optical) are used to determine robust stellar masses and to estimate SFRs and stellar mass growth rates from optical emission lines such as \halpha (\cite[Davidzon \etal\ 2017, Faisst \etal\ 2016a]{DAVIDZON17,FAISST16a}). In addition, deep rest-frame UV spectroscopy from 8-10 meter telescopes (e.g., Keck and VLT) characterize UV emission lines (such as \lya~or \heii) and provide important constraints on metallicity from the measurement of various absorption features (\cite[Ando \etal\ 2007, Faisst \etal\ 2016b]{ANDO07,FAISST16b}).

These large UV to optical surveys lay the foundation for a comprehensive panchromatic study over many orders of frequencies with the newest and next-generation facilities. The \textit{ALPINE} survey with ALMA is a current example of this. The combination of ALMA's sensitivity and efficiency with carefully pre-selected samples from complete optical surveys opens doors for the first large-sample study of dust and gas properties of galaxies in the early growth phase at $z>4$.

\section{\textit{ALPINE}: A Textbook Example of a Panchromatic Study}

The \textit{ALMA Large Program to Investigate \Cii~at Early Times} is a 70-hour ALMA cycle 5 program (\#2017.1.00428.L) to carry out targeted Band-7 observations of $122$ main-sequence galaxies in the early growth phase at $4 < z < 6$ (Figure~\ref{fig:fig1}). 
 \textit{ALPINE} is a diverse, multi-national collaboration of $\sim45$ scientists, led by the PI-team of O. LeF\`evre (PI), A. Faisst (North America science lead), M. B\'ethermin (ALMA data reduction lead), P. Cassata, P. Capak, D. Schaerer, J. Silverman, and L. Yan. As of today (December 2018), $85\%$ of the data has been acquired and reduced. For more information on \textit{ALPINE}, we refer to future survey papers.

\begin{figure}[t!]
\begin{center}
\includegraphics[width=0.99\columnwidth]{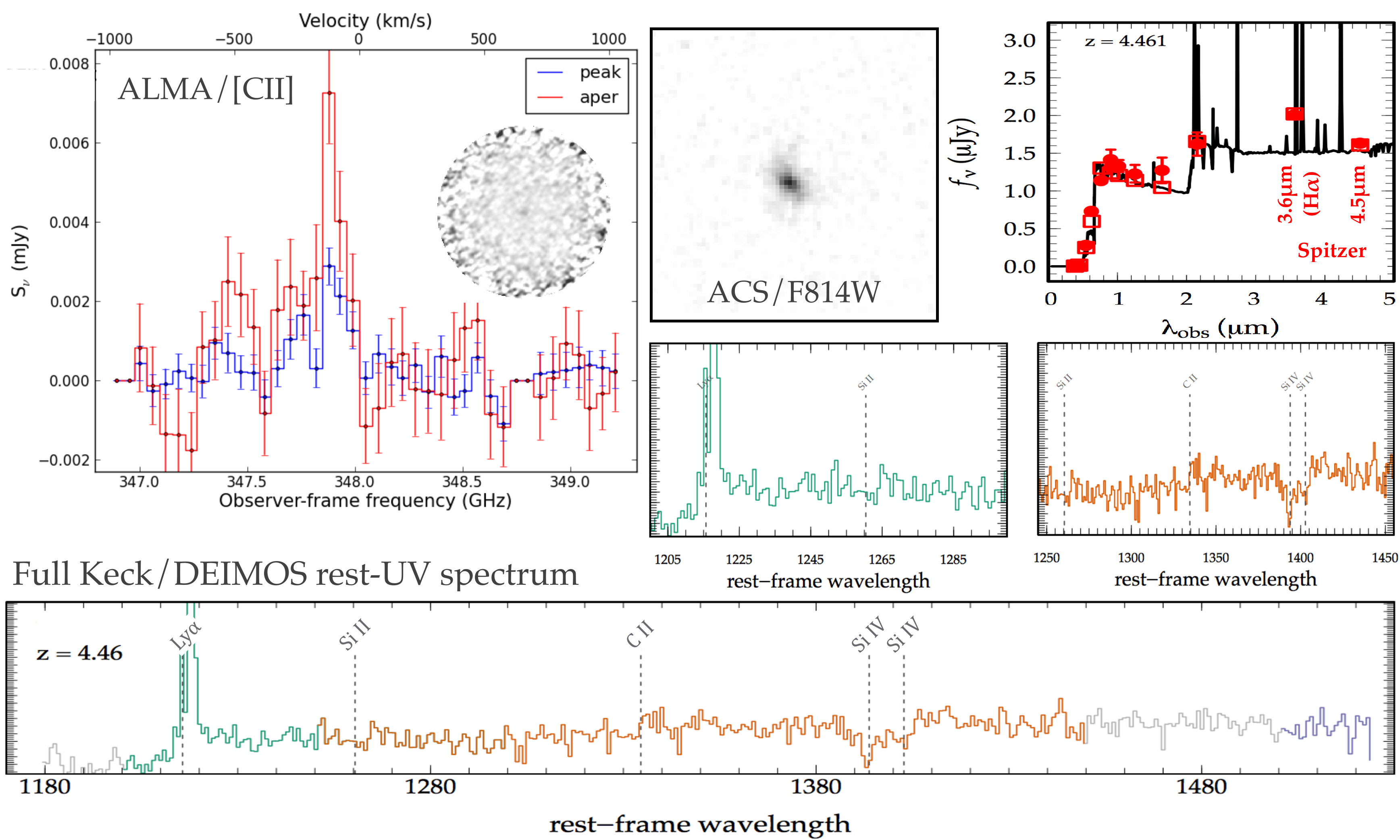} 
\caption{Demonstration of the power of panchromatic data on a galaxy from \textit{ALPINE}. \textit{Clockwise:} \Cii~emission and dust continuum from ALMA; HST imaging; multi-band SED and \halpha from Spitzer color; rest-frame UV spectrum from Keck/DEIMOS (\lya~and absorption lines).}
\label{fig:fig2}
\end{center}
\end{figure}

\textit{ALPINE} maximizes its efficiency and impact by building upon a carefully pre-selected sample in the \textit{Cosmic Evolution Survey} (COSMOS, \cite[Scoville \etal\ 2007]{SCOVILLE07}) and the \textit{Extended Chandra Deep Field South} (E-CDFS, \cite[Giacconi \etal\ 2002]{GIACCONI02}) fields, well studied fields at UV to near-IR wavelengths. The sample is drawn from large spectroscopic surveys undertaken at Keck (\cite[Hasinger \etal\ 2018]{HASINGER18}) and VLT (\cite[LeF\`evre \etal\ 2015]{LEFEVRE2015}), including galaxies with diverse spectroscopic properties in rest-frame UV emission and absorption lines. This approach leads to a complete probe of a diverse galaxy population above ${\rm SFR} \sim 10\,\Msol/{\rm yr}$ in a stellar mass range of $9 < \logm < 11$.  By following up UV-selected galaxies, \textit{ALPINE} probes more than $80\%$ of the total missing flux density at these redshifts, while blind surveys with ALMA cover the remaining less than $20\%$ (Figure~\ref{fig:fig1}).
In combination with its multi-wavelength ancillary data, \textit{ALPINE} is a textbook example of one of the first large panchromatic surveys probing galaxies in the early Universe (Figure~\ref{fig:fig2}).



\subsection{ALPINE Science in a Nutshell}


\textit{ALPINE} builds upon its successful small-scale pre-cursor study based on $10$ galaxies at $z\sim5.5$ (\cite[Riechers \etal\ 2014, Capak \etal\ 2015]{RIECHERS14,CAPAK15}) and reaches a similar sensitivities (continuum: $50\,{\rm \mu Jy}$ or $\log(L_{\rm FIR}/\Lsol) \sim 10.3$; line: ${\rm S_{CII}}\Delta v \sim 0.05\,{\rm Jy\,km/s}$ or $\log(L_{\rm CII}/\Lsol) \sim 7.5$) and resolution ($0.5''-0.7''$).
Its main goal is to quantify the evolution of obscured star-formation, the gas and dust content, and gas kinematics of galaxies in the early universe via \Cii~emission and far-IR continuum measurements at rest-frame $\sim150\,{\rm \mu m}$.
\Cii~is one of the dominant coolants, hence one of the strongest far-IR lines. In addition the atmospheric transmission is high, making these observations favorable.
\textit{ALPINE} combines several science goals and is designed to answer the following questions.

\underline{How does the total star-formation evolve at $4 < z < 6$?}
\textit{ALPINE} provides a comprehensive study of the total SFR density at these redshifts from UV and far-IR continuum measurements. Independent SFRs from far-IR, \Cii, and \halpha (derived from Spitzer colors, \cite[see Faisst \etal\ 2016a]{FAISST16a}) provide the first measurement of the SFR versus stellar mass main-sequence on different timescales of star-formation at these redshifts.

\underline{Is \Cii~a reliable SFR indicator at high-$z$?} The prevalence of \Cii~emission in high-redshift galaxies is a promising tool to estimate SFRs of far-IR continuum faint galaxies. \textit{ALPINE} allows to calibrate this locally well established (\cite[De Looze \etal\ 2014]{DELOOZE14}) relation at $z>4$ where galaxies are more metal poor.

\underline{How dusty are galaxies in the early universe?} Small-sample studies of the \IRXB~relation, which relates the total dust extinction to the line-of-sight dust obscuration, suggest diverse dust properties at $z\sim5$ (\cite[Faisst \etal\ 2017]{FAISST17}). \textit{ALPINE}'s large sample constrains the diversity and studies its causes by a comprehensive analysis of its ancillary data. 

\underline{How does the gas fraction evolve at $z>4$?} \textit{ALPINE} constrains gas masses from far-IR continuum, \Cii, and dynamical mass measurements and measures gas kinematics from \Cii~profiles. It is the first study to trace the evolution of gas mass out to $z=6$.

\textit{ALPINE} unites current UV to far-IR facilities and will in the future build a benchmark sample for various simulations and set the basis for follow-up observations with the James Webb Space Telescope and other next-generation ground- and space-based facilities.

\section{Conclusions}
Large photometric and spectroscopic surveys at UV to near-IR wavelengths provide a solid basis to build a full panchromatic sample over the next couple of years.
Importantly, these samples build on a variety of galaxies, which makes optimally planned targeted follow-up observations with ALMA and later JWST and TMT/E-ELT efficient in covering a significant amount of the parameter space of galaxy properties.
A current example is \textit{ALPINE}, which is specifically built to extend the wavelength coverage of a carefully chosen large spectroscopic sample with ground- and space-based multi-wavelength photometric ancillary data. This makes it to one of the currently largest panchromatic samples to study the formation and evolution of normal main-sequence galaxies in the early Universe. In the future, the \textit{ALPINE} sample will provide a solid basis for follow-up observations with various next-generation facilities.


\end{document}